\documentclass[11pt,a4paper]{emulateapj}
\usepackage{natbib}
\usepackage{amssymb, amsmath, amsbsy}
\usepackage{color}
\usepackage[linkcolor=blue]{hyperref}%

\slugcomment{{\sc Accepted to Astrophysical Journal}}
\usepackage{epsfig}
\usepackage{amsmath}
\usepackage{amssymb}
\usepackage{natbib}
\usepackage{subfigure}
\usepackage{graphicx}
\usepackage{threeparttable}

\newcommand{\smm}{SMM\,J02399$-$0136}
\newcommand{\smmtiny}{SMM\,J02399}

\newcommand\mlya{\rm Ly\alpha}
\newcommand\lya{Ly$\alpha$}
\newcommand\lyb{Ly$\beta$}
\newcommand\mkms{\rm km \, s^{-1}}

\newcommand\mflux{\rm \; erg \, s^{-1} \, cm^{-2}}
\newcommand\mvlya{v_{\rm Ly\alpha}}
\newcommand\vlya{$\mvlya$}
\newcommand\mwlya{W_{\rm Ly\alpha}}

\begin{document}
\def\mean#1{\left< #1 \right>}
\newcommand{\sbunit} {
  erg\ s$^{-1}$ cm$^{-2}$ arcsec$^{-2}$}

\title{Discovery of a \lya\ emitting dark-cloud within $z \sim 2.8$ SMMJ02399$-$0136 system}

\author{
Qiong Li\altaffilmark{1},
Zheng Cai\altaffilmark{2,8},
J.~Xavier Prochaska\altaffilmark{2},
Fabrizio Arrigoni Battaia\altaffilmark{3},
Rob J.~Ivison\altaffilmark{3,4},
Edith Falgarone\altaffilmark{5},
Sebastiano Cantalupo\altaffilmark{6},
Matt Matuszewski\altaffilmark{7},
James Don Neill\altaffilmark{7},
Ran Wang\altaffilmark{1},
Chris Martin\altaffilmark{7},
Anna Moore\altaffilmark{7}
}

\affil{$^1$ Kavli Institute for Astronomy and Astrophysics, Peking University, Beijing 100871, People's Republic of China}
\affil{$^2$ UCO/Lick Observatory, University of California, 1156 High Street, Santa Cruz, CA 95064, USA }
\affil{$^3$ European Southern Observatory, Karl-Schwarzschild-Strasse 2, 85748 Garching, Germany}
\affil{$^4$ Institute for Astronomy, University of Edinburgh, Blackford Hill, Edinburgh EH9 3HJ, UK}
\affil{$^5$ LERMA/LRA, Observatoire de Paris, PSL Research University, CNRS, Sorbonne Universites, UPMC Universite Paris 06, Ecole normale superieure, 75005 Paris, France}
\affil{$^6$ Institute for Astronomy, Department of Physics, ETH Zurich, CH-8093 Zurich, Switzerland}
\affil{$^7$ Cahill Center for Astrophysics, California Institute of Technology, 1216 East California Boulevard, Mail code 278-17, Pasadena, California 91125, USA}
\affil{$^{8}$ Hubble Fellow, zcai@ucolick.org}





\begin{abstract}
We present Keck/KCWI integral field spectrograph observations
of the complex system surrounding
\smm\ (a lensed $z=2.8$ sub-mm galaxy),
including an associated \lya\ nebula,
a dust-obscured, broad-absorption-line quasar, and neighboring
galaxies.   At a 3$\sigma$ surface brightness contour
of 1.6$\times$ 10$^{-17}$ erg s$^{-1}$ cm$^{-2}$ arcsec$^{-2}$,
the \lya\ nebula extends over 17 arcsec ($\gtrsim 140$ physical kpc)
and has a total Ly$\alpha$ luminosity of
$2.5 \times  10^{44} \, \rm erg \, s^{-1}$ (uncorrected for lensing).
The nebula exhibits a kinematic shear of $\sim 1000 \, \mkms$
over 100\,pkpc with lowest velocities east of \smm\ and increasing
to the southwest.
We also discover a bright, \lya\ emitter, separated spatially
and kinematically from the nebula,
at a projected separation of $\approx$60~kpc from the quasar.  This
source has no clear central counterpart in deep {\it Hubble Space Telescope}
imaging, giving an intrinsic
\lya\ rest-frame equivalent width greater than 312~\mbox{\AA} (5-$\sigma$).
We argue that this ``dark cloud'' is illuminated by the quasar with
a UV flux that is orders of magnitude brighter than the emission along
our sightline.
This result confirms statistical inferences that luminous quasars at
$z>2$ emit UV radiation anisotropically.
Future KCWI observations of other lines, e.g.\ \lyb, He\,{\sc ii},
C\,{\sc iv}, etc, and with polarimetry
will further reveal the origin
of the \lya\ nebula and nature of the dark cloud.
\end{abstract}

\keywords{quasars: general ---
galaxies: high-redshift ---
galaxies: formation ---
(galaxies): intergalactic medium ---
cosmology: observations}



\section{Introduction} \label{sec:intro}


In the unified model of AGN emission from \cite{1993ARA&A..31..473A},
the ionizing radiation propagates anisotropically from
an active black hole due to the absorption from a clumpy torus.
For this reason, AGNs appear differently depending
on their inclination with respect to our line of sight.
This model thereby unifies observations of two main classes of AGN:
Type I quasars, which have both broad and narrow emission lines,
and Type II quasars, which have only narrow lines
\citep[e.g.][]{2007A&A...474..837T,2008ApJ...685..147N}.
Both types of AGN are characterized by extreme ionizing luminosities
(up to $\sim 10^{47}$erg s$^{-1}$), which should reveal the
surrounding gas distribution out to large distances,  if illuminated
\citep{1988MNRAS.231P..91R,2001ApJ...556...87H,2005ApJ...628...61C}.

Following this idea, over
the last few years, a few enormous Ly$\alpha$ nebulae (ELANe)
with sizes up to 500 kpc were discovered around
luminous Type I and a few Type II QSOs
at $z=2$--3 (e.g.\ \citealt{2014Natur.506...63C, 2013ApJ...766...58H,2018MNRAS.473.3907A,2017ApJ...837...71C,
2015Sci...348..779H}).
Moreover, deep narrow band (NB) imaging and Integral Field Spectroscopy have revealed
the existence of Lyman-alpha line emitters with high equivalent width ($EW_0>240$\AA) around $z=2$--3 quasars, also known as ``dark galaxies'' (e.g.\
\citealt{2012MNRAS.425.1992C,2017arXiv170903522M}).
Overall, the simultaneous and detailed studies of the kinematics, metallicity, luminosity and abundance of the ELANe and dark galaxies provide us an indispensable opportunity to
understand the interactions between the galaxies hosting AGN at high redshift
and the gas that surrounds and fuels them
(e.g.\ \citealt{2009ApJ...690.1558P,2011MNRAS.418.1796F}).

With these motivations in mind, we have initiated a survey of $z>2$ quasars
and previously known \lya\ nebulae with
the Keck Cosmic Web Imager \citep[KCWI;][]{2012SPIE.8446E..13M},
a new blue-sensitive integral field spectrograph with
wavelength coverage spanning from $\sim 3500-5600$\AA.
Its field-of-view (FoV), spectral resolution, and capability for precise
sky subtraction are optimal for studying gas on scales of $\approx 200$\,kpc.
Here we report on deep KCWI observations of the complex
system surrounding \smm\ \citep{1998MNRAS.298..583I}, hereafter \smmtiny, the first galaxy selected at submm wavelengths (i.e.\ a SMG), gravitationally amplified by the massive foreground galaxy cluster Abell~370.
Typical SMGs are dust-rich and highly obscured. In particular, the Ly$\alpha$ nebula surrounding SMM J02399 was quickly found to lie at z $\sim$ 2.8, using the 4 m Canada-France-Hawaii Telescope, in part because its Ly$\alpha$ luminosity is extremely bright ($\sim 10^{44} \, \rm erg \, s^{-1}$; \citealt{1998MNRAS.298..583I}). This source resides at the high luminosity end of the WISE detected LAB in \cite{2013ApJ...769...91B}, which are also \lya\ nebulae powered by SMGs and have a \lya\ luminosity of $10^{42} - 10^{44} \, \rm erg \, s^{-1}$.
It was the first SMG detected in CO, revealing it to contain a massive reservoir of molecular gas \citep[$\approx 10^{11}$ M$_{\odot}$,][]{1998ApJ...506L...7F}.
This gas reservoir was later shown to cover a large volume and contain multiple, merging galaxies \citep{2010MNRAS.404..198I}, one of which is a dusty, broad-absorption-line (BAL) quasar \citep[][cf.\ \citealt{2003ApJ...584..633G}]{1999A&A...346....7V,1999A&A...351...47V,2001A&A...380..409V,2018ApJ...860...87F}.
The existence of a large reservoir of diffuse molecular gas has been recently inferred by observations of the CH$^+$ cation (Falgarone et al. in prep.), following the detections against several other SMGs \citep{2017Natur.548..430F}.

Inspired by these earlier results,  we observed the system
with KCWI tuned to the \lya\ nebula and directed to cover it
and its neighboring galaxies and quasar.
To our surprise, the data reveal a new, luminous
\lya\ source -- the primary focus of this paper, which is
structured as follows.
We describe our KCWI observation and data reduction process in \S\ref{sec:obs}. In \S\ref{sec:result}, we present the detection of the $\gtrsim100$ kpc Ly$\alpha$ nebula powered by this system and the discovery of a new dark cloud
illuminated by the nearby BAL QSO. Finally, we discuss the Ly$\alpha$ nebula and the physical interpretations of the newly discovered dark galaxy in \S\ref{sec:discussion}, and give a brief summary in \S\ref{sec:Conclusion}.
Throughout this paper, we assume a flat cosmological model with
$\Omega_{\Lambda} = 0.7, \Omega_{m} = 0.3$ and $H_0 = 70 \, \mkms \, \rm Mpc^{-1}$
which implies the physical scale is $\approx 7.85 \, \rm kpc$ per arcsec at
the redshift of \smm.
We follow the naming convention
presented in \cite{2010MNRAS.404..198I}:
L1 marks the BAL quasar; L2 is a low-surface-brightness companion to the East of L1; L2SW is a massive, extremely luminous and dust-obscured
starburst galaxy, East-South-East of L1 and South-West of L2;
L1N is a faint, compact northern component, visible in post-SM2 {\it HST} imaging.

\section{Observations and Data Reduction} \label{sec:obs}

\subsection{KCWI observations}

The KCWI observations of \smmtiny\ were carried out on UT 2017 October 21 (seeing $\sim$ 1.5'')
using KCWI on the Keck II telescope of the W.\,M.~Keck Observatory in Hawaii.
We used the BM1 grating and the medium slicer (slice width $\sim0.7''$) which yields an IFU datacube with FoV of $20'' \times 16.8''$
(pixel scale of $0.3'' \times 0.7''$)
centered on J023951.88$-$013558.0,
the quasar optical position given by \citet{1998MNRAS.298..583I}.
The grating was tilted to give a  central wavelength of 4620\AA\ and
provides a spectral resolution, $R \approx 4000$.
The good wavelength coverage is $\sim$ 4230 - 5010\AA\ .
The total on-source exposure time is four 10-min exposures, each dithered by $\sim 0.6''$.
For sky observations, we used an `offset-target-field' to construct the sky datacube.
The offset-target has a different redshift and is located $\sim 2$ degrees from \smmtiny.
It is a compact point-source at the Ly$\alpha$ wavelength of \smmtiny.

To convert the spectral images and calibration frames (arcs, flats, bias)
to a calibrated datacube, we used the
IDL-based KCWI data reduction
pipeline\footnote{Available at \url{https://github.com/kcwidev/kderp/releases/tag/v0.6.0}}.
Basic CCD reduction is performed on each science frame
to obtain a bias-subtracted, cosmic-ray-cleaned and gain-corrected image.
The continuum flat images are employed for CCD response corrections and pixel-to-pixel variations.
We used a continuum-bar image and an arc image (ThAr)
to define the geometric transformations and wavelength calibration,
generating a rectified object data cube (see the pipeline documents\footnote
{\url{https://github.com/kcwidev/kderp/blob/master/AAAREADME}}).
Twilight flats were used for slice-to-slice flux correction,
and the data was corrected for atmospheric refraction.
Each object and sky frame was flux calibrated with the standard star, Hiltner\,600.
For the sky frame, we first masked
the point-source of the offset-target, and then estimated the sky level at each wavelength channel by
the median of unmasked sky pixels. Then, for each channel, we subtract the sky
from \smmtiny.
For each exposure
we found the QSO centroid to measure the
offsets between exposures, and then performed a weighted mean
with inverse-square variance weighting to construct the final data cube.

\subsection{Ancillary Data}
CO(3--2) observations of \smmtiny\ were obtained by \cite{2003ApJ...584..633G} with a
synthesized beam of $5.2 \times 2.4$ arcsec$^2$.
We use these data, kindly provided by L.~Tacconi, to compare the cool molecular gas emission with that
traced by \lya.
\smmtiny\ is strongly magnified by Abell~370, which was imaged as part of
the {\it HST} Frontier field
survey (HST-14038, PI: J.~Lotz).
It was observed with three broadband filters:
ACS/WFC F435W, F606W and F814W, for total exposure times of 12, 6 and 29~hr, respectively.
We retrieved the publicly available reduced data\footnote{ \url{http://archive.stsci.edu/}}
and produced an average image, weighted by the exposure time.
The F435W filter covers the Ly$\alpha$
emission of \smmtiny\ and thereby
provides an independent estimate of its spatial extent.

\begin{figure*}
\centering
\includegraphics[height=5in]{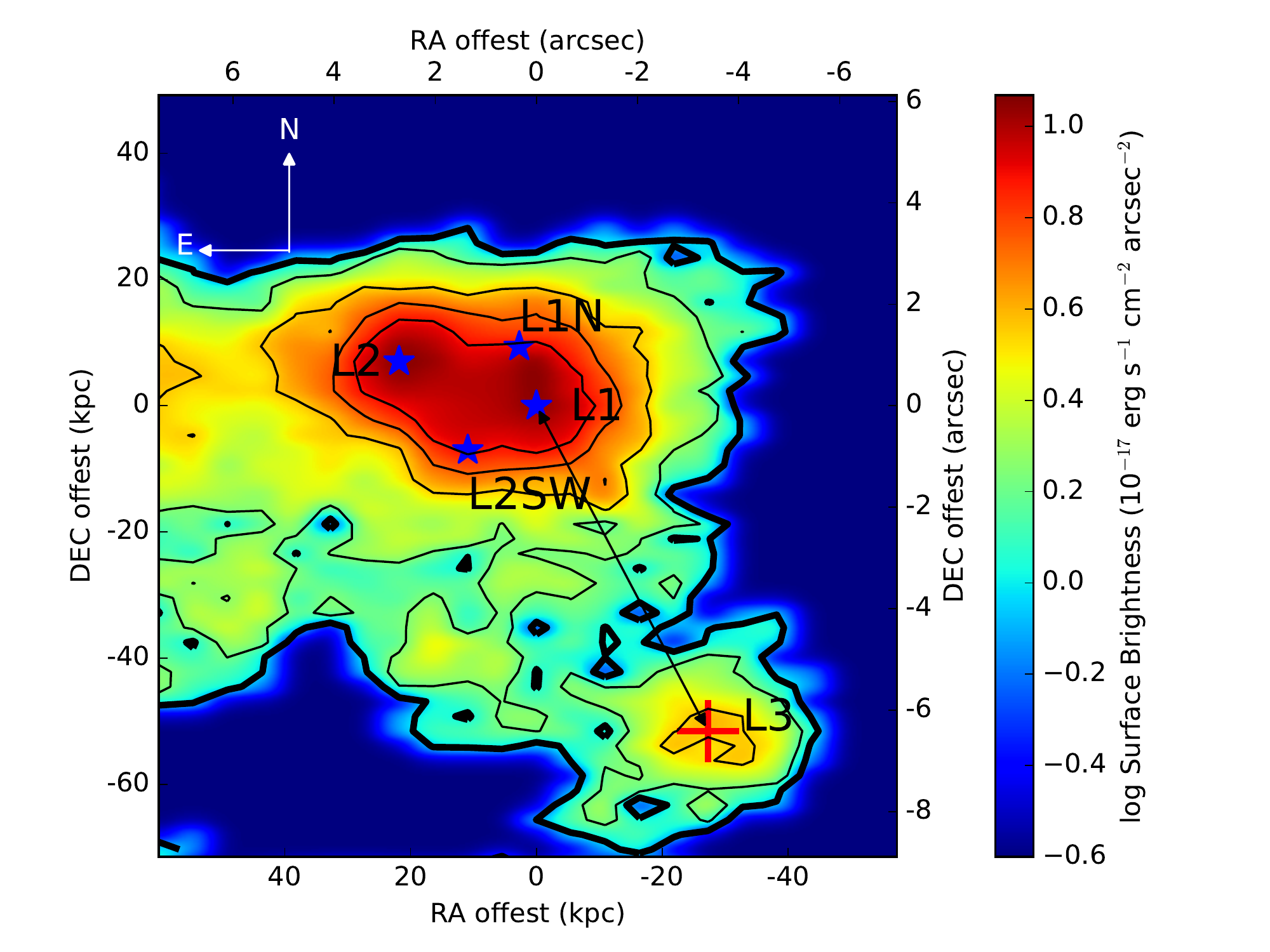}
\caption{The continuum subtracted pseudo-narrow band Ly$\alpha$ image of the gas surrounding
\smmtiny. The color map and the contours indicate the Ly$\alpha$ flux density and the signal-to-noise ratio (+3, 5, 10, 15, 20$\sigma$), respectively.
The axes are centered on the position of L1
and the image has been smoothed with a Gaussian kernel of $2''$.
The offset from L1 to L3 is 3.45" W, 6.45" S ($\sim$58.8kpc).
}
\label{fig:NB_image}
\end{figure*}
\section{Results} \label{sec:result}
\subsection{Deep Narrow-band Imaging at \lya}\label{sec:figure1}

Fig.~\ref{fig:NB_image} shows our
continuum-subtracted, pseudo-narrow band image, spanning
$\approx  30$\AA\ with central wavelength
$\lambda_{\rm NB} = \lambda_{\rm Ly\alpha} (1+z_{L1}) \approx 4025 $\AA, and Gaussian smoothed spatially by $2''$.
The $3\sigma$ depth flux density of this image in a 1.7 arcsec$^2$ aperture
\footnote{In this paper, we used two kinds of apertures. One ($\sim1.7 \rm \, arcsec^2; 1.4\arcsec \times 1.2\arcsec$) matches the seeing of the observations and it is used to subtract spectral features and measure the continuum in the HST image. The second ($\sim 6 \rm \, arcsec^2$) is defined by the $5\sigma$ contour at the L3 location (see Fig.~1) and is used to measure its total/average properties. This second aperture is used to calculate the observed (Section 3.1 para.2) and inferred results (in Section 4.1).}
is $4 \times  10^{-18}  \rm erg \, s^{-1} cm^{-2}$\AA$^{-1}$, corresponding to a surface brightness of $1.6\times 10^{-17} \rm erg \, s^{-1} cm^{-2} arcsec^{-2}$.
In the following discussion,
we refer only to observed quantities (i.e.\ without lensing corrections
for magnification by Abell~370 \footnote{The magnification factor of L1, L2, L3 are almost the same ($\sim$2.4$\times$ for each). \citep{2010MNRAS.404..198I}}.)
The peak surface brightness in the region encompassing L1 and L2
is $\approx 1.2 \times 10^{-16} \rm \, erg \, s^{-1} \, $ $\rm cm^{-2} \, arcsec^{-2}$.
To a 3$\sigma$ surface brightness contour, this nebula has a lensing-uncorrected size
of $\gtrsim 17$ arcsec (i.e.\ $\gtrsim 140$ physical kpc) and
has a total Ly$\alpha$ luminosity of
$2.5 \times  10^{44} \, \rm erg \, s^{-1}$ .

Fig.~\ref{fig:NB_image} also reveals a bright, previously
unreported source approximately $7.5''$ to the SW of L1.
We refer to this source as L3 and measure a
\lya\ luminosity of $1.3 \times 10^{43} \rm \, erg\,s^{-1}$
within an aperture of 6 arcsec$^2$.
The peak surface brightness of L3
is $3.0 \times 10^{-17} \rm \, erg \, s^{-1} \, cm^{-2} \, arcsec^{-2}$,
similar to \lya\
blobs surveyed by \citet{2004AJ....128..569M,2011MNRAS.410L..13M},
as well as the extended \lya\ nebulae detected
around $z \sim 2$ high-redshift radio galaxies and radio-loud quasars (e.g.\ \citealt{2007MNRAS.378..416V,1991ApJ...381..373H}).

\begin{figure*}
\centering
\includegraphics[width = \linewidth]{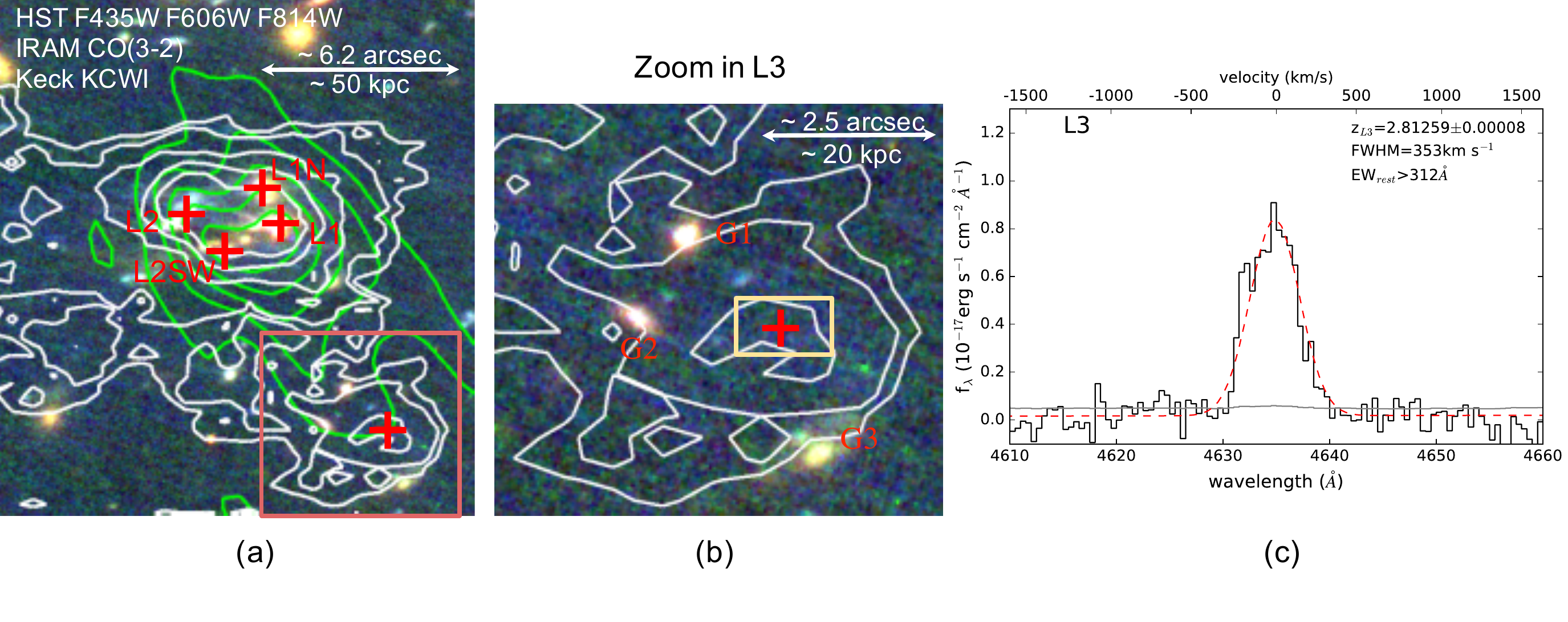}
\caption{Left: A false color {\it HST} image of \smmtiny\, from
the Frontier field survey (HST-14038, PI: J. Lotz).
We overlay on it KCWI narrow band contours (white) and CO J=3-2 contours (green, \citealt{2003ApJ...584..633G}). Middle: zoom in on
the bright \lya\ emitter (L3) marked by red box in the
left panel.
Right: Ly$\alpha$ emission of L3 through the
aperture of $\sim$1.7 arcsec$^2$ ($1.4'' \times 1.2''$, yellow box in Middle panel).
The red dashed line shows a Gaussian fit to the emission profile.
The gray line indicates the noise spectrum.}
\label{figure2}
\end{figure*}

Fig.~\ref{figure2}a shows the {\it HST} imaging around L3,
comprised of ACS/WFC F435W, F606W and F814W deep images and
contours for the CO and \lya\ emission.
CO(1-0) and CO(3-2) observations suggest there is a large gas reservoir in this system, and the peak location of which coincides with the \lya\ emission at L1.
We have also confirmed that there is no continuum (stellar) counterparts corresponding
to L3 in the SDSS, WISE and 2MASS catalogs
\footnote {SDSS catalog: \citet{2015ApJS..219...12A,2009ApJS..182..543A} \\
WISE catalog: \citet{2012yCat.2311....0C,2013yCat.2328....0C} \\
2MASS catalog: \citet{2003yCat.2246....0C,2003yCat.7233....0S}}
, nor in the deeper surveys of PanSTARRS, DES, DECaLS.
\footnote{The Pan-STARRS1 Surveys: \citet{2016arXiv161205560C} \\
DES DR1: \citet{2018ApJS..239...18A} \\
DECaLS: \url{http://legacysurvey.org/dr7/}}
Examining Fig.~\ref{figure2}b, we
derive a
$5 \sigma$ upper limit to the continuum
$4.0 \pm 0.8 \times 10^{-20} \mflux$~\AA$^{-1}$ from the
{\it HST} F435W image in a 1.7 arcsec$^2$ aperture.
The \lya\ rest-frame equivalent width is therefore
greater than 312\AA,
which follows the scenario of a dark galaxy\footnote{The definition of `dark galaxy' is defined as a Ly$\alpha$ emitter with rest-frame EW greater than 240\AA\ , following the definition of \citet{2012MNRAS.425.1992C}.}
illuminated by a luminous AGN
\citep{2012MNRAS.425.1992C}.

Here we also estimated the $5 \sigma$ continuum upper limits from both F606W and F814W imaging in an aperture of 1.7 arcsec$^2$, obtaining $1.9 \times 10^{-20} \mflux$~\AA$^{-1}$ and $9.7 \times 10^{-21} \mflux$~\AA$^{-1}$, respectively. Then we calculated that the $5 \sigma$ upper limit of the rest-frame UV SFR is $0.3 \, \rm yr^{-1}$ \citep{2012ARA&A..50..531K}. This SFR is not high enough to give rise to the bright Lyman-alpha emission.

\begin{figure*}
\centering
\includegraphics[width = \linewidth]{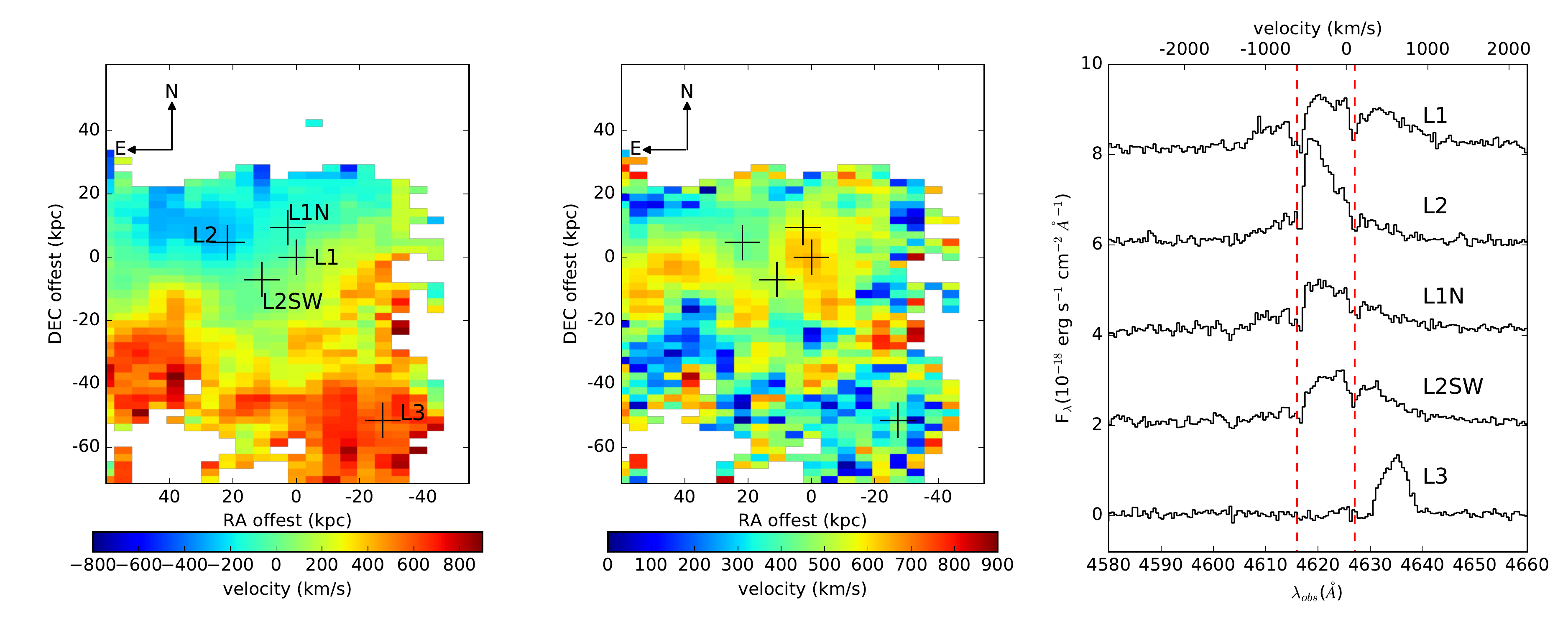}
\caption{Left: flux-weighted velocity-shift map with respect to the systemic redshift of L1, with 3$\times$2 pixels (1.2''$\times$1.4'') in a top-hat filter.
Middle: velocity dispersion map obtained from the second-moment of the flux distribution.
Right: spectra at several locations in
the system, observed through an aperture of $1.2'' \times 1.4''$. The spectra
are shifted by $2 \times 10^{-18} \, \rm erg \, s^{-1} \, cm^{-2} \, \AA^{-1}$
for presentation purposes.
The vertical red lines are a coherent absorption system. }
\label{figure3}
\end{figure*}

\begin{table*}
\caption{The properties for individual source in SMMJ02399}
\begin{center}
\begin{threeparttable}
\vspace{0.2cm}
\begin{tabular}{lccccccc}
\hline \hline \noalign {\smallskip}
 Source&RA &DEC &$\lambda_{\rm detected}$ &redshift &Aperture &SB$(Ly\alpha)$&L$(Ly\alpha)^a$    \\
         & J2000      &  J2000         &$\rm \mathring{A}$    &   &        & $\rm 10^{-17} erg s^{-1} cm^{-2} arcsec^{-2}$   &$\rm 10^{43} erg s^{-1}$    \\
   (1)   &  (2)  &  (3)      & (4)       &  (5)  & (6)     &  (7)  &(8)     \\
\hline \noalign {\smallskip}
L1         &02:39:51.86   &-01:35:58.15     &4625.5   &2.8048$\pm$0.0004  &$\sim$0.7''$\times$0.6''    &8.3$\pm$1.8 &3.5$\pm$0.6     \\
L2         &02:39:52.04   &-01:35:57.27     &4621.6   &2.7985$\pm$0.0004  &$\sim$0.7''$\times$0.6''    &8.3$\pm$1.7 &3.5$\pm$0.6        \\
L3         &02:39:51.63   &-01:36:04.56     &4634.9   &2.81259$\pm$0.00008 &$\sim$1.4''$\times$1.2''   &3.0$\pm$0.6 &1.3$\pm$0.2  \\
Entire nebula&            &        &  &&&&25.0$\pm$3.6\\
\hline \noalign {\smallskip}
\end{tabular}
\small \textbf{Notes.} \\
a. the observed luminosities, which do not include the lensing magnification factor of 2.4$\times$ due to the foreground cluster.\citep{2010MNRAS.404..198I}

\end{threeparttable}
\label{source}
\end{center}
\end{table*}

\subsection{Ly$\alpha$ Kinematics}\label{sec:figure3}

Our KCWI datacube enables a study of \lya\ emission
throughout the complex system surrounding \smmtiny.
Fig.~\ref{figure3} shows
the flux-weighted centroid velocity \vlya\
relative to $z_{\rm L1}$ and the velocity dispersion $\sigma_v$.
A gradient in \vlya\ is evident as one traverses
from L2 to L3 through the nebulae ranging from $-300$ to $+800 \, \mkms$.
In contrast, the velocity dispersion is roughly constant
at $\sigma_v \approx 500 \mkms$.

We extract 1D spectra from the final datacube for L1, L2, L1N, L2SW and L3 through a aperture of 1.7 arcsec$^2$
(Fig.~\ref{figure2}c and Fig.~\ref{figure3}c).
The \lya\ emission of L3 is well-modeled by a single Gaussian
with central wavelength
$\lambda_{\rm L3}^{\rm Ly\alpha} = 4634.9 \pm  0.1$\AA\ corresponding
to $z_{\rm L3} = 2.81259 \pm 0.00008$.
The Gaussian model also gives a
a velocity dispersion of
$\sigma_{\rm L3}^{\rm Ly\alpha} = 150 \pm 6 \mkms$
(or FWHM$_{L3}$ = $353\pm 15 \mkms$).

In Fig.~\ref{figure3}c, except for L3, the \lya\ emission from the nebula
shows both
broad and narrow components, confirming
the previous long-slit spectra \citep[see also][]{2001A&A...380..409V,2010MNRAS.404..198I}.
The nebular emission cannot
be well described by a single Gaussian.
The figure also emphasizes that L3 is kinematically distinct
from the larger nebula;  its centroid is offset by
several hundred $\mkms$ and it exhibits no broad component.

The \lya\ emission spectrum of the nebula also shows
evidence for two weak absorption features on both the
red and blue sides of the primary emission.
Intriguingly,
line observations of the CH$^+$ cation, a specific tracer of turbulent
dissipation \citep{2017Natur.548..430F},
reveal broad emission lines ascribed to shocks
and  a broad absorption line (FWHM = 600 $\mkms$)
in the direction of L2SW
(Falgarone et al., in prep.).
\lya\ and CH$^+$ emission of this system
both show broad line-emission ascribed to shocks and narrower
lines that may imply that L1 and L2 are encompassed
by a large and massive HI cloud.

\section{Discussion} \label{sec:discussion}

\subsection{Illumination of the dark cloud (L3)}\label{sec:Dark galaxy}

\begin{figure*}
\centering
\includegraphics[width = \linewidth]{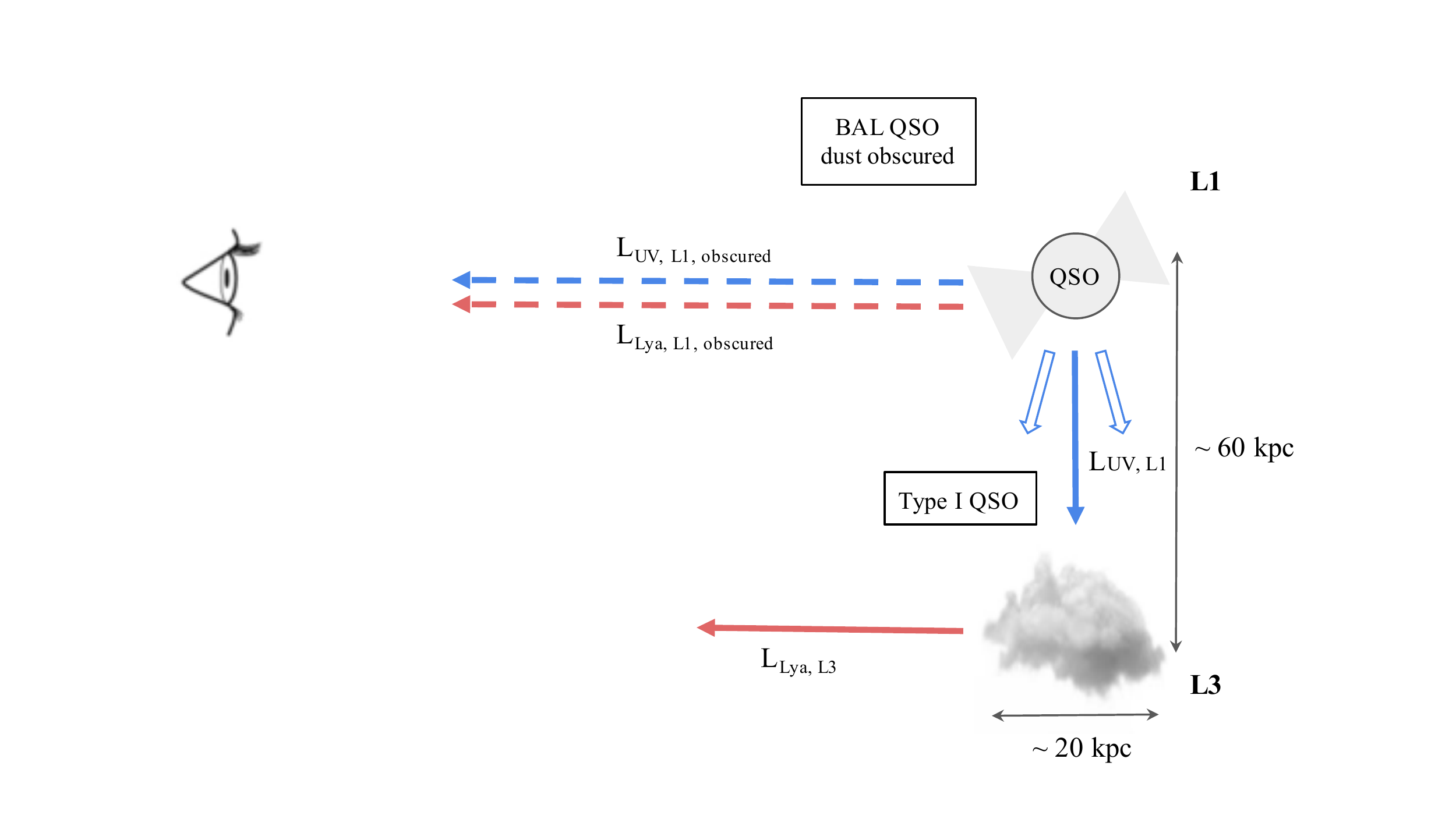}
\caption{A schematic diagram of the aniostropic emission of L1, illuminating the
neighboring, and otherwise dark cloud, L3.}
\label{figure4}
\end{figure*}

In the previous section, we reported on the discovery of L3, a luminous
\lya\ emitter with an extremely high \lya\ equivalent width
($\mwlya >312$\AA) and a FWHM of $\approx 350 \mkms$.
\cite{2012MNRAS.425.1992C} reported on a sample of sources
with $\mwlya > 240$\AA\ surrounding the ultra-luminous quasar
HE0109$-$3518 which lack detectable continuum counterparts in deep broad-band imaging.
They termed these sources
``dark galaxies'' and argued they were illuminated by the quasar
as no star-formation could power such high $W_{\rm Ly\alpha}$,
i.e.\ the observed \lya\ is fluorescent radiation powered by
the incident, ionizing photons.
The properties of L3 are consistent with this dark-galaxy scenario,
except that the observed UV emission from L1 is orders of
magnitude lower than HE0109$-$3518.
Therefore, we propose a scenario wherein the dust-obscured BAL
quasar is emitting UV radiation anisotropically,
with bright UV emission emitted transverse to our line of sight
(Fig.~\ref{figure4}).


We may test the \lya\ fluorescence hypothesis by comparing the observed
\lya\ flux with that predicted for optically thick gas illuminated
by the {\it unobscured} ionizing radiation of L1.
To estimate the unobscured luminosity of L1,
consider the following analysis.
L1 is detected by the WISE satellite with a $W2$ magnitude
at $\approx 4.6 \mu$m of $15.2\pm0.1$\,mag from ALLWISE source catalog.
Adopting the Type-I QSO template from \cite{2006ApJS..166..470R} and
scaling to this $W2$ measurement we may estimate $F_{\nu,LL}$,
with $\nu_{LL}$ the frequency at the HI Lyman limit.
We estimate
$L_{\nu,LL}$ of L1 to be $\log_{10} (L_{\nu,LL} / \rm erg \, s^{-1} \, Hz^{-1}) = 30.07$
and note that this model also reproduces the observed $W1$ flux.
In contrast, we estimate the obscured UV luminosity at $\nu_{LL}$
along our sightline to be only
$\log_{10} (L_{\nu,LL} / \rm erg \, s^{-1} \, Hz^{-1}) = 27.74$,
using the Type-II QSO template from \cite{2007ApJ...663...81P}
and the QSO ultraviolet template from \cite{2015MNRAS.449.4204L}.

We further assume:
(1) the redshift offset between L1 and L3 is due to peculiar motions
rather than the Hubble flow, such that the physical separation
between the two is comparable to the projected offset;
(2) L3 is an optically thick gas cloud with radius
of $R_{\rm L3} \approx 10 \, \rm kpc$,
estimated from the NB image (Fig.~\ref{fig:NB_image});
(3) the optically thick cloud converts a fraction
$\eta_{\rm thick} =0.66$ of the ionizing photons from the quasar to \lya\
photons emitted at a uniform brightness \citep{1996ApJ...468..462G,2013ApJ...766...58H};
(4) L3 is `half-moon' illuminated with a geometric reduction
factor of $f_{\rm gm} =0.5$ consistent with radiative transfer simulations \citep{2005ApJ...628...61C,2010ApJ...708.1048K}.

From these assertions, we derive the \lya\ surface
brightness \citep{2013ApJ...766...58H},
\begin{equation}\label{FIR}
\begin{split}
{ SB }_{\mlya}
&=\frac { f_{gm} \eta_{\rm thick} h \nu_{Ly \alpha}  }{ \left( 1 + z  \right)^4 } \frac {\Phi}{\pi}   \\
& = 5.1 \times 10^{-17} \left( \frac{1 + z}{3.8}  \right)^{-4} \left( \frac{f_{gm}}{0.5}  \right)
\left( \frac{R}{60 \, \rm kpc}  \right)^{-2} \\
&\times  \left( \frac{L_{\nu, LL}}{1.2 \times 10^{30} \, \rm erg \, s^{-1} \, Hz^{-1}}  \right)
  {\rm erg \, s^{-1} \, cm^{-2} \, arcsec^{-2}}
\end{split}
\end{equation}
where $\Phi$ (phot s$^{-1}$ cm$^{-2}$ ) is the ionizing photon
number flux, and $R$ is the distance of L3 from L1.
Integrating this surface brightness over the observed size of L3,
we estimate $f_{\rm L3}^{Ly\alpha} = 3.2 \times 10^{-16} \, \rm erg \, s^{-1} \, cm^{-2}$.
The observed flux of \lya\ from L3 is $f_{\rm L3,obs}^{Ly\alpha} = 1.8 \times 10^{-16} \, \rm erg \, s^{-1} \, cm^{-2}$, corresponding to surface brightness of $SB_{\rm L3,obs}^{Ly\alpha} = 3.0 \times 10^{-17} \rm erg \, s^{-1} \, cm^{-2} \, arcsec^{-2}$.
This derived value is consistent with the observed flux of \lya\ from L3.

Adopting fluorescence as the origin of the \lya\ emission,
we further estimate a cool gas mass
of $M_{\rm cool} \sim  1.7 \times 10^{10} M_{\odot}$ using
Equation 8 in \cite{2012MNRAS.425.1992C}, assuming a
clumping factor C = 1 and T = 2$\times 10 ^{4}$K.
We have also used CLOUDY ionization modeling
\citep{1996hbic.book.....F} with the same assumptions above, and
found that the best parameter combinations to match L$_{Ly \alpha, L3}$ is
($ \rm N_{H} , Z , n_H $)$\sim$ ($10^{22} \rm cm^{-2} , 0.01 Z_{\odot} ,  0.1 cm^{-3}$).
Future metal line observations may provide more precise limits.
Assumed H density of L3 is $\sim$ 0.1 cm$^{-3}$ according to our best CLOUDY model, the collapse timescale t$_{ff}$ is $\sim$ 1.6 $\times$ 10$^8$ years.
Considered the cosmic correction, t$_{ff}$ from earth is $\sim$ 6.1 $\times$ 10$^8$ years, $\sim$ 4.5\% of the universe age.
And considered `dark matter', t$_{ff}$ will be shorter.
The integrated opacity due to Thompson scattering of the ionized cloud is negligible given its very low cross-section ($0.665 \times 10^{-24} \rm cm^{2}$) even given our relatively large total electron column density of $ \rm N_{e} \sim 10^{22} \rm cm^{-2}$.  Even a very luminous quasar will give an insignificant flux compared with what we observed.

Numerical studies indicate that the dark galaxy phase corresponds
to gas-rich galaxies prior to efficient star formation
(e.g.\ \citealt{2005ApJ...628...61C,2012MNRAS.425.1992C,2010ApJ...708.1048K,2017arXiv170903522M,2004MNRAS.353..301F,2009ApJ...693L..49H}).  A few dedicated
dark galaxy surveys using deep narrow-band imaging
and the Multi Unit Spectroscopic Explorer (MUSE) (e.g.\ \citealt{2012MNRAS.425.1992C},\citealt{2017arXiv170903522M}) found that the dark clouds reside in
the mass range M$_{\rm gas} \sim 10^{\rm 8-10}$ M$_{\odot}$, and in the
\lya\ luminosity range of $10^{\rm 41-43}$ erg s$^{-1}$.
The dark cloud discovered here is consistent with these properties.

Note that there are three galaxies within $5\arcsec \times 5\arcsec$ from L3 in the deep HST images, which we indicate in panel b of Fig. 2 as G1, G2, and G3.
The broadband colors of G1 and G3 are consistent with Lyman break galaxies at $z=3$,
though with large uncertainties given the faintness of these galaxies.
Even if we allow that G1 and G3 are partially powering L3 and adopt a UV flux from those sources, then the EW$_{\rm Ly\alpha}$ estimated for L3 decreases to 209 - 270 \AA.
Nevertheless, the Ly$\alpha$ kinematics of L3 has a FWHM of $353\pm 15 \mkms$ which is relatively quiescent. Also, the emission peak has a large spatial offset($\sim$2.0'') from both galaxies. These evidences support that L3 is not likely to be powered by either G1 or G3.

Indeed, we cannot rule out a scenario where L3 is powered by gravitational infall, but our current data supports that L3 is more likely to be externally illuminated. Future deep polarimetric observations may rule out such scenarios.



\subsection{\lya\ tracing anisotropic QSO emission}\label{sec:LAB}

The dimensions, luminosity, and kinematics of the \lya\ nebula surrounding \smmtiny\ are reminiscent of the MAMMOTH-1 nebula which extends to 450~kpc \citep{2017ApJ...837...71C}
and is also powered by a Type-II AGN \citep{2018A&A...620A.202A}.
\cite{2017ApJ...837...71C} argued that the powering mechanism of MAMMOTH-1
could be a combination of photoionization and
shocks due to an AGN outflow (e.g.\ \citealt{2014MNRAS.441.3306H}).
\smmtiny\ is another nebula hosting a highly obscured system,  and
may have a similar powering mechanism.
For the full nebula around \smmtiny\, we measure a total Ly$\alpha$ luminosity of
$2.5 \times 10^{44}$ erg s$^{-1}$ (uncorrected for lensing).
The AGN bolometric luminosities is
$2.5 \times 10^{46}$ erg s$^{-1}$ estimated by the UV
luminosities (1450$\mathring{A}$) with $L_{bol} = 4.2\nu L_{\nu},1450 \mathring{A}$ \citep{2012MNRAS.422..478R}.
Considered the lensing magnification, the BAL QSO has a similar bolometric luminosity ($\sim 10^{46}$ erg s$^{-1}$) with a normal QSO at z=3.
To compare with MAMMOTH-1, we subtracted the Ly$\alpha$ PSF.
The PSF is constructed using a full-width-half-maximum that is equal to the seeing of $1.5''$. The PSF amplitude is determined from the Ly$\alpha$ luminosity of L1 (in an aperture of $\sim1.7$ arcsec$^2$) convolved with a Moffat kernel.
We find that the central PSF luminosity constitutes only 4.4\% of the entire nebula, which is similar to the MAMMOTH-1 measurement of 4\%.
In the scenario involving collimation, previous studies have already shown that L1 may have a jet/outflow betrayed by the radio morphology, and L2 is likely explained as the shock-excited region or a reflection nebula (\citealt{2010MNRAS.404..198I,2018ApJ...860...87F}). High-resolution
CO observations with ALMA and JVLA, and red-sensitive observations of extended and broad
metal-line emission (e.g.\ C$_\textrm{\scriptsize IV}$) \citep{2017ApJ...837...71C},
could provide decisive evidence for the outflow scenario.

We note that this Ly$\alpha$ nebula is not the only one which contains multiple embedded galaxies and lacks a clear continuum sources at the peak of the Ly$\alpha$ extended emission.
Currently, $\sim10$ Ly$\alpha$ nebulae (also known as Ly$\alpha$ blobs, LAB) are reported as powered by obscured sources. Examples of these are the LABs in \cite{2012ApJ...752...86P}, \cite{2013ApJ...769...91B}, and a few LAB in the SSA22 field (e.g.\ \citealt{2004AJ....128..569M,2007ApJ...667..667M}). With an extent of $>$140 kpc and a luminosity of $2.5 \times 10^{46}$ erg s$^{-1}$, our nebula is one of the largest and most extended among these systems, and intriguingly hosts the most luminous powering source.
Regarding the previous samples of Ly$\alpha$ nebulae and LABs, the SMM J02399 nebula is one of the most luminous discovered to date, and also has multiple components. Also, this is the only system containing a proto-galaxy (a dark galaxy by the definition of \citealt{2012MNRAS.425.1992C}), indicating a highly asymmetric ionization.

Fig.~\ref{figure4} illustrates
the anisotropic emission of ionizing radiation from L1 that we envisage with
L1 unobscured in the direction of L3.
The UV photons ionize the outer layers of the
cool gas cloud to produce Ly$\alpha$ emission.
Along our line of sight it is a typical dust obscured BAL quasar with
substantial foreground absorption,
supporting unified models of AGN.
This confirms the primary conclusion from a
series of papers that have studied the anisotropic clustering
of optically thick gas transverse to and along the line of sight to quasars
\citep[e.g.]{2007ApJ...655..735H,2013ApJ...766...58H}.
Based on the same assumption in \S \ref{sec:Dark galaxy},
here we estimate a lower limit of the open angle of L1 conservatively as $\Omega / 4\pi > 0.007$
($f_{obscured} = 1 - \Omega / 4\pi < 0.993$), with the half-angle $\theta > 9.5^\circ$. The opening angel is estimated using the linear size of L3 and the distance between L1 and L3, assuming L1 and L3 is at the same redshift, and that the redshift offset is due to the kinematics not Hubble flow.
Further evidence supporting the anisotropic emission hypothesis is the BAL nature of L1, which has been constrained by the moderate continuum polarization with VLT/FORS1 \citep{2001A&A...380..409V}. BAL QSOs are expected to radiate as Type-I QSOs only through patches free of dust and dense outflowing gas (e.g., \citealt{1999ApJS..125....1O}).
On the other hand, the SMG (L2SW) close to L1 also can be a plausible source to illuminate L3, which dust cocoon may not necessarily be homogenous.



\section{Concluding Remarks}
\label{sec:Conclusion}

In this paper, we present the Keck/KCWI IFU observations of a Ly$\alpha$
blob powered by \smmtiny\ at z $\sim$ 2.8. With KCWI, we discover a dark cloud
that we argue is illuminated by the dust-obscured QSO of the system.
This implies strong, anisotropic UV radiation from the QSO, which was
also suggested by previous polarimetry observations.
The future red/blue sensitive IFS, Keck Cosmic Reionization Mapper (KCRM)/KCWI can further reveal the properties and kinematics of other lines, such as Ly$\beta$, He$\textrm{\scriptsize II}$, C$\textrm{\scriptsize IV}$, to further
reveal the nature of this and similar systems.
We are also pursuing a KCWI survey of Type-II AGN to
study the population of \lya\ nebulae in a set of
fully obscured sources.

{\bf Acknowledgement:}
This work was supported by National Key Program for Science and Technology
Research and Development (grant 2016YFA0400703) and the National Science Foundation of China (11721303). We are thankful for the supports from the National
Science Foundation of China (NSFC) grants No.11373008, 11533001.
QL gratefully acknowledge financial support from China Scholarship Council.
ZC acknowledge the supports provided by NASA through
the Hubble Fellowship grant HST-HF2-51370 awarded by the
Space Telescope Science Institute.
, which is operated by the Association of Universities for Research in Astronomy, Inc., for NASA, under contract NAS 5-26555.
EF gratefully acknowledges funding from the European Research Council, under the European Community’s Seventh framework Programme, through the Advanced Grant MIST (FP7/2017-2022, No 787813).
The data presented herein were obtained at the W. M. Keck Observatory, which
is operated as a scientific partnership among the California Institute of Technology, the University of California and the National Aeronautics and Space Administration. The Observatory
was made possible by the generous financial support of the W. M. Keck Foundation.


\end{document}